# Evanescent light-matter Interactions in Atomic Cladding Wave Guides

Liron Stern[1], Boris Desiatov[1], Ilya Goykhman[1], and Uriel Levy[1]

[1]Department of Applied Physics, The Benin School of Engineering and Computer Science, The Center for Nanoscience and Nanotechnology, The Hebrew University of Jerusalem, Jerusalem, 91904, Israel
*Corresponding author: ulevy@cc.huji.ac.il

**Alkali vapors, and in particular rubidium, are being used extensively in several important fields of research such as slow and stored light[1,2] non-linear optics[3] and quantum computation[4,5]. Additionally, the technology of alkali vapors plays a major role in realizing myriad industrial applications including for example atomic clocks[6,7] magentometers[8] and optical frequency stabilization[9]. Lately, there is a growing effort towards miniaturizing traditional centimeter-size alkali vapor cells. Owing to the significant reduction in device dimensions, light matter interactions are greatly enhanced, enabling new functionalities due to the low power threshold needed for non-linear interactions. Here, taking advantage of the mature Complimentary Metal-Oxide-Semiconductor (CMOS) compatible platform of silicon photonics, we construct an efficient and flexible platform for tailored light vapor interactions on a chip. Specifically, we demonstrate light matter interactions in an atomic cladding wave guide (ACWG), consisting of CMOS compatible silicon nitride nano wave-guide core with a Rubidium (Rb) vapor cladding. We observe the highly efficient interaction of the electromagnetic guided mode with the thermal Rb cladding. The nature of such interactions is explained by a model which predicts the transmission spectrum of the system taking into account Doppler and transit time broadening. We show, that due to the high confinement of the optical mode (with a mode area of $0.3\lambda^2$), the Rb absorption saturates at powers in the nW regime.**

Light matter interactions with alkali vapors are the subject of flourishing research field for several decades. In addition to providing basic information on the properties of these atoms, such interactions hold a grand promise for variety of applications spanning from quantum memories[5], magnetometry[8] and few photon switching[3] to the control of speed of light by the slow and fast light effects[1,2]. Currently, alkali vapors such as Rb and Cesium (Cs) are already serving as key components in modern navigation and communication[6,7]. Following the general trend for miniaturization and high level integration, there is an ongoing effort towards tightly confining light and vapor in order to obtain efficient, miniature and low-power light-vapor interactions. Recent works demonstrated significant reduction in the interaction volume by the use of micro and nano vapor cells[10-12]. In parallel, several works demonstrated the interaction of thermal alkali vapors with light in guided mode configuration. This includes the use of hollow core photonic crystal fibers[13-16] (HC-PCF), hollow core anti reflecting optical wave-guides[17-19] (HC-ARROW), tapered nano-fibers[20-22] (TNF) and nano wave-guides[23]. In an HC-PCF, alkali vapor is pumped into the core of the fiber and interacts with the electromagnetic field which is guided along the core of the HC-PCF. This platform served for the demonstration of variety of effects including vapor spectroscopy[13,15], electromagnetic induced transparency[14] (EIT), and enhanced two-photon absorption[16]. Alternatively, the TNF approach, consisting of solid core fiber surrounded by Rb vapor cladding, was already shown to be useful for nano-watt saturation of the atomic transition[20], two photon absorption[21] and efficient all optical switching[22]. A further step towards miniaturization and integration was accomplished by constructing hollow core waveguides on a chip, via the use of the HC-ARROW[18,19] approach. This work has the benefit of being both compact, chip-integrable and enabling strong light matter interaction on chip. It has been further demonstrated the ability to achieve EIT and consequently slow light in this platform[19].

In spite of the fact that the HC-ARROW concept provides an important step towards miniaturization and integration, reducing the dimensions of the guided mode towards the sub micron regime remains an immense challenge. In this work we tackle this challenge by establishing an on chip platform for strong light matter interactions in a chip scale environment by using a solid core waveguide surrounded by Rb vapor cladding. This configuration can be considered as the on chip equivalent of the TNF approach, where the evanescent tail of the electromagnetic mode interacts with the alkali vapor. A natural choice of platform for the realization of the solid core waveguide in the submicron regime is that of the silicon photonics, which enables strong light confinement and dense integration of devices on a chip. Indeed, the emerging field of silicon photonics is becoming a key technology in the construction and the demonstration of various nanoscale devices for myriad applications such as communication, signal processing, and sensing to name a few[24,25]. Unfortunately silicon core waveguides cannot be used for the interaction of light with Rb or Cs atoms because of their lack of transparency in the relevant spectral window between 770-900 nm. Yet, one can use silicon nitride (SiN) as the waveguide core material. In addition to being used routinely in the CMOS industry, SiN offers several prominent advantages; it is transparent both in the visible and the infrared spectrum, it possesses moderately high refractive index (n~2) thus allowing tight confinement of the optical mode. Additionally, being a dielectric material with large energy bandgap SiN does not suffer from free carrier's absorption at wavelengths above 300nm. Another advantage of SiN is its low thermo-optic coefficient (~$10^{-5}$ $K^{-1}$) making it less sensitive to environmental variations. Following these properties, sub micron SiN waveguides were recently demonstrated for sensing, filtering and nonlinear optics applications [26-29].

Here we demonstrate an atomic cladding wave guide (ACWG) consisting of a SiN core nano wave-guides integrated with an alkali vapor cladding, and reveal evanescent light-matter interactions on a chip, with nanoscale dimensions and low light level interactions in both the linear and the nonlinear regime. Using this approach, one can integrate atomic vapor on top of an existing CMOS compatible photonic circuit, and selectively enable the interaction of the light and atoms in arbitrary regions of the chip. For example, one can think of obtaining strong light mater interactions of Rb atoms with nanophotonics resonators, e.g. micro ring/disk resonators, photonic crystal resonators as well as with on-chip nanofocusing devices[30,31] and nanoantennas[32]. Such a platform may also be used for achieving strong-coupling cavity quantum electro-dynamics in a chip-scale environment[33]. Using this concept of ACWG we also observed non linear phenomena in ultra-low levels of light. This is because of the tight confinement of light in our ACWG, having a mode area of $0.3\lambda^2$. Parallel to our experimental demonstration, we suggest a framework to analytically predict the absorption spectra of such ACWGs. Our model includes the specific broadening mechanism unique to our system, i.e. the increased Doppler broadening and evanescent transit time broadening. The model prediction is compared to our experimental results and found to be in good agreement.

Our ACWG is illustrated in figure 1a, where a photonic chip consisting of few SiN wave guides with a silicon oxide cladding is sketched. Above a chosen region on the chip the upper oxide cladding is etched out and a cylinder is epoxy bonded to the chip. This opening will be used as the interaction region between the optical mode guided by the SiN waveguide and the Rb atoms. Next, the cell is connected to a vacuum system, baked out, and a droplet of natural abundance Rb is inserted. In all other regions expect this interaction region, the wave guides are buried under the oxide layer. The fabrication method of such an integrated ACWG is described in the methods.

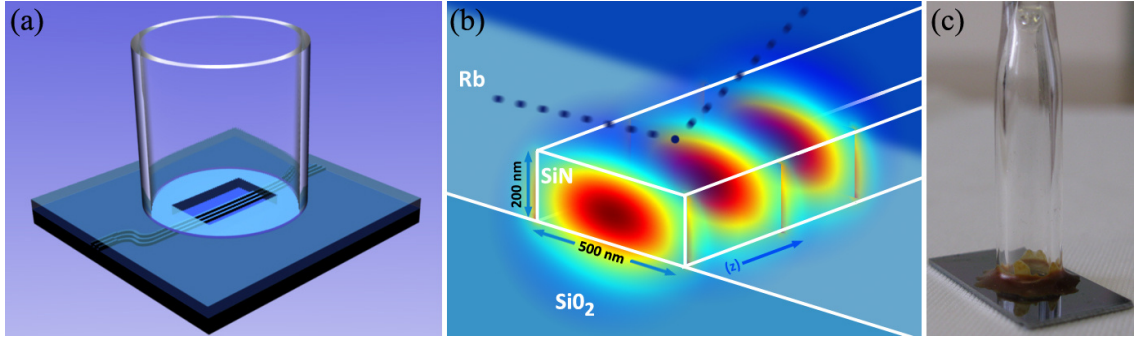

**Figure 1:** (a) Illustration of the ACWG (b) Cross section of the silicon nitride waveguide in the interaction area with the simulated fundamental TE mode superimposed. A Rb atom is illustrated as reflecting the surface, leaving the surface at a quenched state (c) photograph of the device showing the bonding of the vapor cell to the chip.

The wave-guides structure was designed to support a TE (Transverse electric) like mode, in-plane polarization with width and height of 500 nm and 200 nm, respectively. The waveguide dimensions were chosen as a compromise between the desire for strong light-matter interactions and dense integration, requiring high confinement of the optical mode in the waveguide core and the need for significant interaction of light with the surrounding atoms, requiring a substantial portion of the optical mode to reside in the cladding. In addition to affecting the strength of interaction, the specific geometry of the wave-guide is also expected to determine the absorbance profile due to the Doppler broadening, and transit time broadening which are affected by the propagation constant of the optical mode, and the evanescent decay lengths respectively. In Fig. 1b we present an illustration of such a wave guide, with the simulated mode profile calculated by the finite element method (Comsol mutliphysics) superimposed on it. The confinement factor of the mode, i.e. the fraction of electric field that interacts with the Rb, is found to be 10.9%, calculated by integrating the fraction of intensity in the interaction region divided by the unit input power. A Rb atom is illustrated as ballistically striking and reflecting from the surface (dark blue dashed lines). The nature of this process in context of the absorption spectrum will be evident in the next paragraph. In Fig. 1c a photograph of the integrated device is shown.

In order to calculate the effective index of refraction of the atomic cladding layer, and to shed light on the dynamics of the transmission spectrum in our system, we adapt the method developed in the context of attenuated total internal reflection (ATIR). The reflection of light from an interface between a dielectric and a vapor has been investigated thoroughly both theoretically and experimentally, and particularly for the case of total internal reflection[34-37]. Due to the non-local nature of the response of the atoms to the applied field, resulting from the transient response of group of atoms emanating from the surface[34], there is a need to find an effective susceptibility which takes into account the dynamic nature of light matter interactions as a consequence of the collision of the atoms with the wall. While calculating the effective susceptibility one must consider the finite time of interaction. Due to the evanescent decay of field in the vapor, the time of interaction between the atoms and the electromagnetic wave is finite, and is typically lower than the lifetime of the excited level. This effect results in broadening of the spectrum, known as the transit time broadening. Additionally, one must consider the effect of the pseudo-momentum carried by the mode along the propagation direction. This momentum results in Doppler broadening of the absorption spectrum.

In order to find the effective susceptibility one solves the optical Bloch equations, with a boundary condition of total dephasing (quenching) when an atom hits the dielectric. The solution is separated for the steady state solution, for the atoms moving toward the wall (transverse velocity $v_t < 0$) and a transient solution for the atoms moving from the wall (transverse velocity $v_t > 0$). Using these two solutions, and the concept of the fictitious polarization, it can be shown[34,36,37] that the first order susceptibility is given by the following expression:

$$\chi(\nu) = \frac{e^2}{\pi\varepsilon_0 m_e} \sum_{i=1,2} C_i \sum_{F,F'} \frac{Nf_{FF'}}{2(2I+1)\nu_{F_i',F_i}} \int_{-\infty}^{\infty} dv_z \int_0^{\infty} dv_t \frac{W(v_z, v_t)}{-2\pi\Delta\nu - k_z v_z - i(\gamma - ik_t v_t)} \quad (1)$$

Where, N is the density of atoms, e the electric charge of an electron, $\varepsilon_0$ the vacuum permittivity, $m_e$ is the electron mass, $C_i$ is the occupation ratio, $\nu_{Fi'Fi'}$ is the transition frequency from a level with total angular momentum F to an excited state with momentum F', $v_z$ and $v_t$ are the atomic velocities in the propagation direction $z$, and the transverse direction respectively, $W$ is the normalized velocity distribution, $k_z$ and $k_t$ represent the momentum of the evanescent field along the propagation direction and the transverse direction respectively. $2\gamma$ is the natural full-width half maximum line-width and $\Delta\nu$ the frequency detuning from the transition frequency. The oscillator strengths $f_{FF'}$ are defined using the different quantum numbers of the different transitions, as described in supplementary 1.

Eq. 1 is valid in the case of low vapor density, as in our experiments. In the case of higher densities, the formalism should be adapted, in the same manner that has been done in the context of single reflection[37]. Thus in order to calculate the effective refractive index, besides the density of atoms (equivalent to temperature), we are in need to find the momentum in the direction of propagation, and the momentum in the direction perpendicular to the propagation in order to obtain the exact line shape. The first will specify the amount of Doppler broadening, and the later will specify the amount of transit time broadening. In wave-guide terminology these momenta are the effective wave-number, $\beta$ and the inverse evanescent decay length $\eta$, respectively. The higher confined the mode, the higher the $\beta$ and the smaller the $\eta$ resulting in an increase in both Doppler and transit time broadened line-shape. As a result of the broadening, the contrast of the absorption lineshape is expected to decrease. Further reduction in the contrast is also expected because of the reduced interaction of the electromagnetic field with the atomic vapor, resulted from the higher confinement of the mode in the solid core waveguide rather than in the atomic cladding. In order to evaluate the validity of the proposed model, we first compare it to published results of atomic cladding tapered fibers reported independently in reference 20 and reference 21. In both cases, the model predicts the absorption spectrum adequately. This validity comparison is presented fully in supplementary 1. It should be noted, that one can exploit the framework to more elaborate light-matter interactions, as the ATIR susceptibility has been developed in the case of pump-probe experiments[38], as well as selection reflection in the case of dark resonances[39].

Next, we investigate light propagation in our ACWG. For reference, we first measure the transmission spectrum obtained by illuminating the vapor cell directly, i.e. above the ACWG (Figure 2a). In this figure, one can clearly see the distinct Doppler broadened absorption dips corresponding to both $^{85}$Rb and $^{87}$Rb absorption lines. For reference, we also show the eight oscillator strengths of the atomic transitions of $^{85}$Rb (red vertical lines) and $^{87}$Rb (blue vertical lines) in arbitrary units, where the oscillator strength is proportional to the height of the vertical line. Next we couple light directly to the ACWG and observed the obtained spectrum. The normalized transmission spectra of light propagating along the ACWG is shown in Fig 2b. The results were obtained at cell's temperature of about 135°C, corresponding to a density of $3.9 \cdot 10^{19}$ m$^{-3}$ atoms. The signal has been attenuated in order to avoid saturation of the transitions. As can be clearly seen, in comparison with the bulk transmission the absorption lines are broader. As mentioned earlier, this result is associated with an in an additional Doppler broadening due to the increased momentum in propagation direction, and also with an increase in the transit time broadening resulted by the limited interaction time. This broadening manifests for instance in the total overlap of the $^{87}$Rb F=2 → F'=1 and the $^{87}$Rb F=2 → F'=2 transitions with the $^{85}$Rb F=2→F'=2/3 transition. It should be noted that the data presented here was obtained on course of few days. No evidence of degradation of the operation of the device has been witnessed.

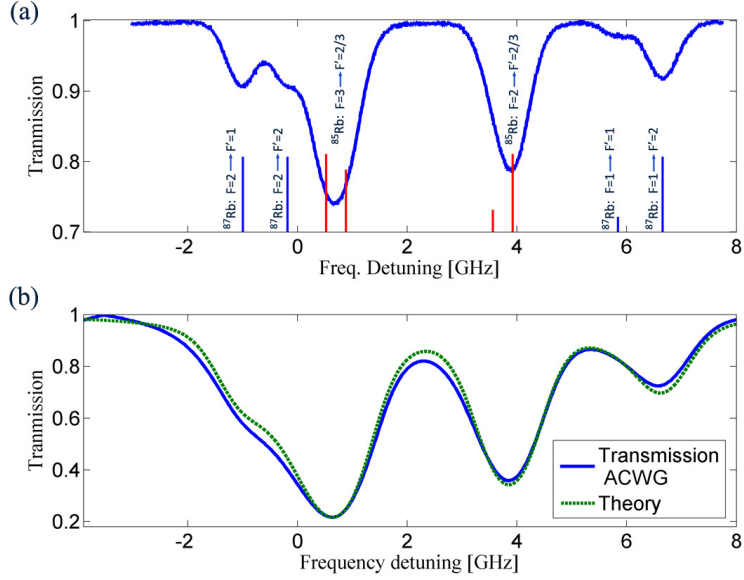

**Fig 2:** (a) Natural Rb Doppler broadened spectrum, as measured through the cell attached to the silicon chip. (b) Normalized measured spectrum (blue) and calculated spectrum (green). Calculations were performed based on the actual dimensions of the ACWG and the density of the vapor.

In order to compare the presented model to our experimental results we plot the simulated transmission spectrum after propagating through our ACWG (Fig. 2b). The spectrum is calculated based on Eq. 1 using the procedure described in supplementary 1. Temperature of $135^0$C, confinement factor of 0.11, interaction length of 1.5 mm, atom density of $3.9 \cdot 10^{19}$ m$^{-3}$, effective index of 1.56 and evanescent wave number of $9.47 \cdot 10^6$ m$^{-1}$ were used to calculate this transmission spectrum. As can be seen, this model predicts both the widths and the contrast of the different absorption lines. The model predicts a broadened line resulting from a Doppler broadening of 900 MHz and transit time broadening of 60 MHz, and an overall width of 1.1 GHz. These values were calculated by assuming no transverse velocity component (for Doppler broadening), no longitudinal velocity component (for transit time broadening), and both velocity components (for the total width).

Next, we investigate nonlinear interactions in our ACWG system. Specifically, we measure the variations in the absorption spectrum as a function of the incident light intensity. To do so, we measure the transmission spectrum with different neutral density (ND) filters set at the output of the laser, whilst the ACWG remains under constant coupling conditions and constant temperature. In Fig 3a we plot the obtained transmission spectrum, whereas in fig 3b we plot the contrast of the F=2 → F'=2/3 transition as a function of the ND filter attenuation ratio. The optical power within the waveguide was estimated by measuring the output power, and taking into consideration the coupling loss in our system. Our current version of the ACWG system is not optimized for maximum coupling efficiency. The overall coupling loss, is currently about 30 dB, including coupling loss from the fiber to the waveguide (2 facets) and from the oxide clad waveguide to the ACWG (two facets). Additionally, the epoxy glue induces some loss. These loss factors will be reduced in the future devices e.g. by increasing the thickness of the oxide clad layer and by optimizing the dimensions of the ACWG with respect to that of the oxide clad waveguide. By assuming a symmetric coupling scenario we estimate the power level within the waveguide to be about 15 dB higher than the power measured at the output.

As can be clearly seen in both Fig. 3a, and Fig. 3b the contrast is increasing with the decrease in the optical power level within the ACWG, as anticipated in a saturated transition scenario. In order to estimate the power where this saturation occurs, we fit the data in Fig 3b to an exponential function of the following form: $T(P) = \exp(-\alpha_0(1+P/P_{sat})^{-1/2}$ Where $\alpha_0$ is the optical density corresponding to the unsaturated level measured, P is the optical power level within the ACWG at resonance (obtained by a linear fit around the resonance) and $P_{sat}$ is the saturation optical power. The square root factor is used as we expect the system to be dominant by in-homogeneous broadening. We find, using these values in Eq. 2 that the estimated onset of saturation is about 40nW. This low power level is consistent with the theoretical and experimental estimations given in [20,21], and is the result of the high confinement of the optical mode compensating both Doppler and transit-time broadening which increase the saturation threshold.

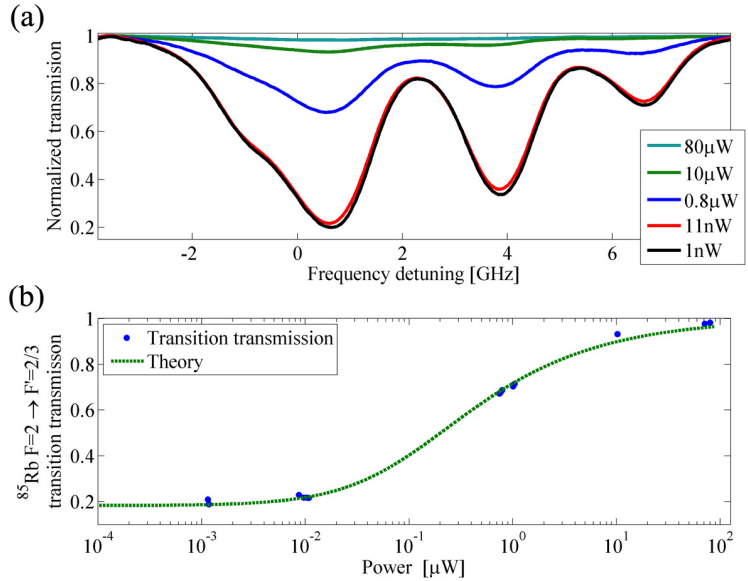

**Fig 3:** Saturation of the Rb transitions. (a) Normalized transmission of the ACWG with different incident power levels, attenuated using neutral density filters (b) Minimal transmission of the F=2 to F'=2/3 $^{85}$Rb transition as a function of the optical power within the ACWG.

Next we investigated the effect of temperature on the transmission of light, while the optical transitions are not saturated. Fig. 4a shows the transmission spectrum at several different temperatures. As can be noticed, the contrast increases with the increase in temperature. This is expected because higher temperatures correspond to higher density of atoms, i.e. stronger absorption. In Fig. 4b the calculated (Using Eq. 1) and the measured contrast of the F=2 to F'=2/3 $^{85}$Rb transition as a function of temperature are shown. Both curves are in good agreement. It is worth noticing, that the absorption is achieved by the interaction of a relatively small number of atoms and photons. Calculating the amount of atoms that interact with the optical mode, at any instant of time can be performed by the multiplication of the evanescent volume and the density of atoms. Such calculation indicates that only a few hundred of atoms are interacting with the photons, at the $100^0$C temperature range and a few thousand atoms at the $130^0$C temperature range.

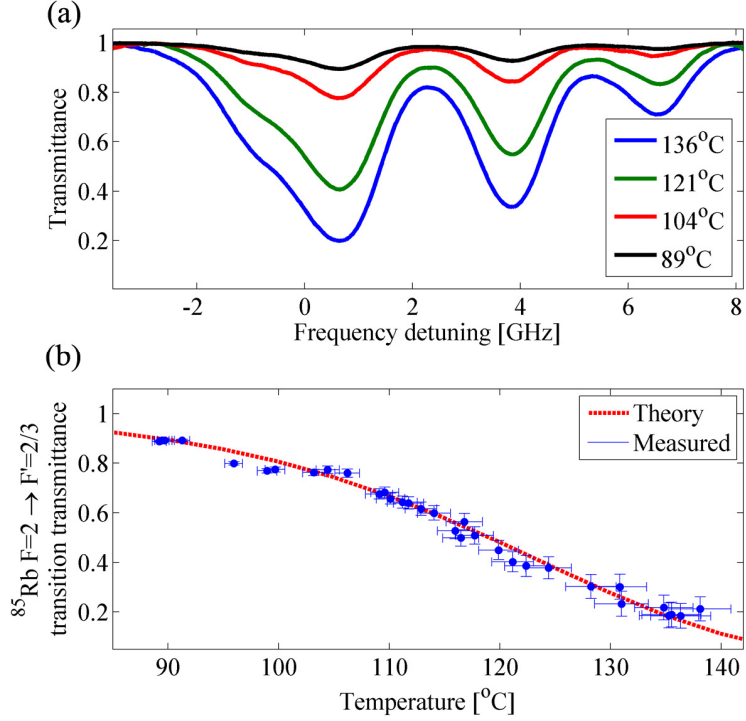

**Fig 4:** Temperature dependence of the absorption. (a) Normalized transmission of the ACWG at different temperatures (b) Transmission of the F=2 to F'=2/3 $^{85}$Rb transition as a function of temperature, and the expected theoretical contrast using Eq. 1.

To summarize, we have developed and presented a chip scale platform providing efficient evanescent light-matter interactions in an atomic cladding wave guide (ACWG). Based on this platform we demonstrated the interaction of an on chip nano scale guided optical mode with Rb atoms in both the linear and the nonlinear regime. Specifically, we show the transmission spectrum of light passing through our device at various optical power levels and various temperatures. From the obtained results we have estimated the saturation to be in the nW regime. This relatively low value is the result of the high mode confinement compensating both Doppler and transit-time broadening which increase the saturation threshold. Additionally, based on the ATIR formalism which also takes into account Doppler and transit time broadening we have developed a model which allows calculating the transmission spectrum in our ACWG structures.

The platform of ACWG offers several prominent advantages; Dimension-wise, an ACWG, has a mode area that is about two orders of magnitude smaller than previously demonstrated on chip approaches[17]. In addition, due to the ability to integrate large variety of existing photonic circuits with the ACWG, this approach can further enhance light-matter interaction via the use of resonators such as micro-ring resonators, and structures such as slot wave guides making this platform attractive for non-linear optics. The ACWG silicon nitride platform further enables to couple broadband optical signals, making it applicable (simultaneously) to various alkali vapors such as Cs and Rb and various optical transitions whether the D1 and D2 lines, or higher excited Rydberg states[40]. Similarly to other systems that exhibit strong light matter interactions with atomic vapor, the ACWG is limited in the obtainable absorption line widths due to finite time of interaction, and the increased Doppler broadening. The later can be reduced using various Doppler-free excitation schemes, resulting in a transit time broadened line, which can be viewed as an effective homogeneous broadening limiting the systems coherence. This for instance will determine the modulation frequency in switching applications[22,41], and the stability of an optical frequency standard and may be decreased using a buffer gas[22]. Nevertheless, due to the unprecedented ability to control the electro-mangetic field distribution in nano-scaled structures, this approach makes possible to conduct controlled and efficient light-matter interactions at the nano-scale, which are very challenging to be achieved in other configurations. These exciting opportunities manifest in both well established and new applications such as few photon switching, quantum gates and logic, slow and fast light and optical frequency standards all in a chip-scale foot-print. Finally, while this paper deals with hot atoms, the demonstrated platform may also be implemented in applications involving cold atoms, e.g. cavity QED, and exploiting the chip for optical traps of atoms taking advantages of the strong gradient of the evanescent electromagnetic field[42-44].

**Methods:**

**Fabrication:**
To construct the device we used silicon substrate with thermally grown 2.5μm thick silicon oxide and low-pressure chemical vapor deposited (LPCVD) 200nm thick silicon-nitride (SiN) device layer atop (Rogue Valley Microdevices). The SiN waveguides were defined using standard electron-beam lithography (Raith e-line 150) with 20 KV acceleration voltage and ZEP-520A as a positive tone electron-beam resist, following by inductively coupled reactive ion etching ICP RIE (Oxford Plasmalab 100) with a CHF3/O2 gas mixture to transfer the pattern into the SiN device layer. Next to waveguide fabrication, the optical structure was covered with 1.5μm silicon oxide layer using plasma enhanced chemical vapor deposition (PECVD). In order to realize an atomic cell, we have carried out an additional lithographic step to determine a chamber area where PECVD oxide layer was later on etched down with buffered hydrofluoric acid (BHF) to expose the waveguides structure to interaction with rubidium vapor. Finally, a Pyrex cylinder was bonded to the chip using thermally cured epoxy. The cell is connected to a turbo vacuum system, and baked out simultaneously. Following bake-out the device has reached a vacuum level of ~$10^{-8}$ Torr. At this stage natural Rb is launched into the cell, and the cell is disconnected and sealed.

**Setup:**
A 795 nm DFB (Toptica DL100) source is fiber coupled to a singled mode polarization maintaining lensed fiber. The polarization direction is set to be aligned with the in plane (TE) polarization mode of the ACWG. The light is coupled from the lensed fiber into the waveguide in a butt coupling configuration. Symmetrically, light is coupled out of the wave-guide into a second lens fiber using a butt coupling configuration and the intensity is measured using a photo-detector (New-Port 2151). The intensity of the launched light is controlled by inserting neutral density (ND) filters at the output of the laser. In addition, a beam splitter is introduced in the output of the laser in order to be able to measure a reference cell or the bulk of the cell attached to the wave-guide simultaneously to the wave-guide measurement. In order to achieve sufficient density of atoms, the device is heated using resistor heaters, and the temperature is monitored using a thermistor. The thermistor is located a few millimeters from the top of the cell. In order to reduce the accumulation of Rb on the surface of the wave guides we maintain a temperature gradient between the top of the cylinder and the surface of the chip. This way, the Rb droplet is created at the top of the cylinder, where the temperature is lower. Thermal simulations indicate a temperature gradient of the order of $10^0$C between the location of the thermisotr and the cold spot in the cell. The cold spot temperature defines the vapor pressure, and consequently the atomic density.

**Acknowledgments**

The authors thank Avinoam Stern, Yefim Barash and Benny Levy from Accubeat Ltd. for the preparation of rubidium cells, and the use of the vacuum facilities, and Selim M. Shahriar for fruitful discussions. I. Goykhman and B. Desiatov acknowledge the Eshkol fellowship from the ministry science and technology. The waveguides were fabricated at the Center for Nanoscience and Nanotechnology, The Hebrew University of Jerusalem.

## Supplementary 1 – Theoretical model and fitting procedure

In order to evaluate the validity of the proposed model presented in the main section, we first compare it to the results of atomic cladding tapered nano fibers (TNF's) published independently in[1] and in[2]. In this section we describe the procedure of fitting the model presented in the letter, to measured results. First, we simulate (using Comsol multi physics) the eigen-mode and field distributions of Silicon dioxide TNFs with diameters of 350nm and 400nm, corresponding to the dimensions reported in [1] and [2] respectively, from which we obtain the effective indices of refraction of the fundamental mode. By fitting the simulated field to an exponentially decaying function we obtain the evanescent decay length. Using these wave-numbers we numerically evaluate the susceptibility presented in Eq. 1, using $^{87}$Rb and $^{85}$Rb oscillator strengths and the resonance frequency data, that can be found for instance in Ref [3]. The Oscillator strengths are computed using the following relation:

$$f_{FF'} = fW^2(J'F'JF;I1)(2F'+1)(2F+1)$$

Where *f* is the D1 or D2 oscillator strength (0.342 and 0.695 respectively), W is the Racah coefficient, F, F' J and J' are the total angular momentum and angular momentum respectively and I is the nuclear spin. We assumed the temperature to be around $100^0$C as reported by [1] and [2], as an initial parameter. Next, we use the local field relation n= $([1+3\chi]/[3-\chi])^{1/2}$ in order to obtain the complex index of refraction, and use it as the complex and dispersive cladding index of refraction. Using this cladding value we simulate the complex effective index of refraction of the fiber mode (β). Finally, we use this effective index in order to propagate the mode in the fiber, over a length L determined by the interaction region. This is done by using the Beer-Lambert relation, namely T=exp(-2βL) In general, the natural width γ has been taken to be 5.9 MHz. To account for other broadening mechanisms, such as self broadening in dense vapor one may use higher values of γ. We note, that the transmission data that was adapted from Ref [1] and [2] has been obtained when the transitions where saturated. We take this saturation into account by introducing a saturation factor for an inhomogeneous broadened medium, i.e. $(1+I/I_{sat})^{-1/2}$. The powers and saturation powers are taken from Refs [1,2] (10nW, 8nW for ref [1] and 5nW, 30nW for ref [2]).

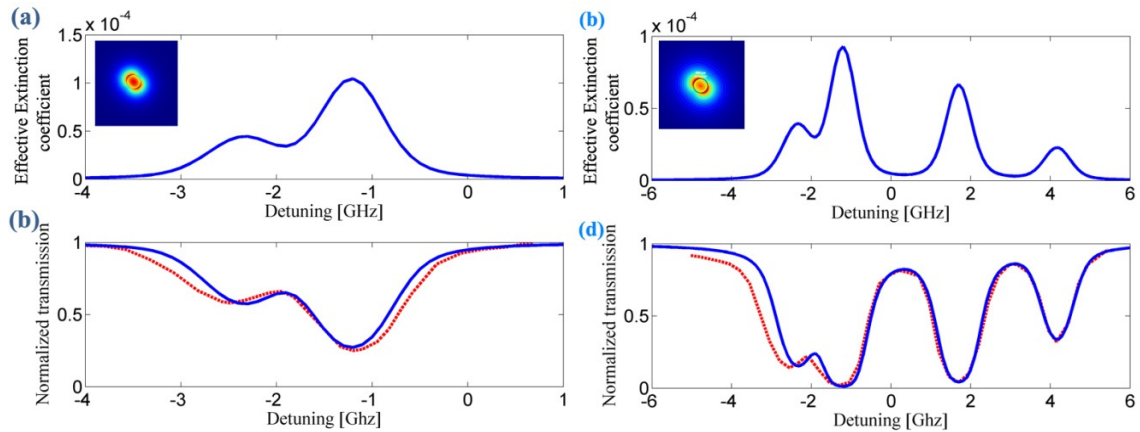

**Fig S1:** (a) Calculated imaginary index of refraction obtained using data from Ref [1] (b) Normalized measured transmission and theoretical transmission in the case of ref [1] (c) Calculated imaginary index of refraction obtained using data from Ref [2] (c) (d) Measured and theoretical transmission in the case of ref [2]

In Fig S1 we show the data of our model compared to the two above mentioned published results. The full set of parameters used in the model is presented in table S1. Fig S1a shows the predicted imaginary part of the refractive index of the atomic cladding of Ref. [1], whereas in Fig S1b the measured (taken from

Ref. 1) and the corresponding calculated transmission spectra (blue and red lines respectively) are plotted. Fig S1c shows the predicted imaginary part of the refractive index of the atomic cladding of Ref. 2, whereas in Fig S1d the measured (taken from Ref. 2) and the corresponding calculated transmission spectra (blue and red lines respectively) are plotted. A good agreement between the model and the measured results is evident. As shown, the model describes both the width and the contrast of the spectrum adequately. The discrepancies between the fitted data and the model can be explained as follows: a) In order to present data in normalized and frequency calibrated units, a linear calibration is usually applied. Errors in this calibration of the order of 5% can explain the discrepancy for instance in Fig S1d. b) The model doesn't fully account to saturation effects, and although we introduced saturation by multiplying the absorption by a saturation factor, a comprehensive treatment of saturation may provide even more accurate results. Such a treatment can take into account for instance the different saturation levels of each transition and the power broadening, all in the framework of a quenched and finite time of interaction, and is beyond the scope of this paper. Nevertheless, the model predicts adequately the amount of transit time and Doppler broadening in both cases. We note, that Eq. 1 was developed in Cartesian coordinates, rather the Cylindrical coordinates. We compensate for this fact by approximating the evanescent tail to an exponentially decaying function rather than to a Bessel function.

Following is a list of Different experimental and simulated parameters used to fit the data presented in Fig S1. The data been retrieved from the published data in the references [1,2]. The only free parameter is the density of atoms. We have found that the equivalent cold-spot temperature is close to that reported by [1] and [2].

| Value: | Ref [1] | Ref [2] |
|---|---|---|
| Temperature | $95^0$C | $88^0$C |
| Confinement Factor | 0.58 | 0.74 |
| Saturation factor | 0.71 | 0.92 |
| Interaction Length | 3 mm | 5 mm |
| Effective Index | 1.1246 | 1.0760 |
| Evanescent Wave Number | 5.89 $10^6$ [m$^{-1}$] | 4.83 $10^6$ [m$^{-1}$] |

**Table S1:** The experimental and the simulated parameters used to fit the data presented in Fig S1